\documentclass[aps,twocolumn,showpacs]{revtex4}
\usepackage{epsf}

\begin{document}
\voffset 0.5 true in
\title{Probing neutron star superfluidity with gravitational-wave data}

\author{N.~Andersson$^{1}$ and G.~L.~Comer$^{2}$}

\affiliation{$^1$ Department of Mathematics, University of Southampton, 
Southampton SO17 1BJ, United Kingdom \\$^2$ Department of Physics, Saint 
Louis University, St.~Louis, MO 63156-0907, USA}  

\begin{abstract}
We discuss the possibility that future gravitational-wave detectors
may be able to detect various modes of oscillation of old, cold 
neutron stars. We argue that such detections would provide 
unique insights into the superfluid nature of neutron star cores, 
and could also lead to a much improved understanding of pulsar glitches.
Our estimates are based on a detector configuration with several
narrowbanded (cryogenic) interferometers operating as a ``xylophone'' 
which could lead to high sensitivity at high frequencies.  
We also draw on recent advances in our understanding of the dynamics
of pulsating superfluid neutron star cores.
\end{abstract}


\maketitle

{\em The extremes of physics}. --- Since their discovery in the 
late 1960s neutron stars have emerged as unique probes of
many extremes of physics.  Comprising roughly one and a half solar 
masses of material  compressed inside a radius of ten kilometers, 
i.e well beyond nuclear density, the neutron stars still hide  
many of their mysteries. For example, we do not yet fully 
understand the nature of the supranuclear equation of state required
to describe matter in the core of neutron stars.
Still, a picture has emerged in which superfluidity 
plays a vital role.  Theoretically, such a picture has as its 
foundation \cite{sauls} the extraordinarily successful many body 
theory of Fermi liquids and the BCS mechanism that are used to describe 
superconductors and superfluid Helium three. It is now generally
believed that once a neutron star cools below a 
few times $10^9$~K (a few months after its birth) the bulk of its
core will become superfluid. Thus, the more than 1000 observed 
pulsars provide useful laboratories for studying 
large scale superfluidity and truly high $T_c$ superconductors.
 
The main observational evidence for neutron star superfluidity
is provided by the glitches (a sudden spin-up followed by a 
long term relaxation period) that have been observed in roughly
30 pulsars. For a summary of the current results see Table~5 in
\cite{wang00}.  While the smaller glitches in, for example, 
the Crab pulsar
can be understood in terms of spindown induced
quakes in the neutron star's crust \cite{franco00}, this model cannot 
explain the large glitches 
seen in, for example, the Vela pulsar \cite{alpar81}.
In the long favoured model for large glitches the  
observed spin-up is caused
by the transfer of angular momentum from a superfluid
that coexists with lattice nuclei
in the inner crust (extending from neutron drip to
$\sim 2\times 10^{14} {\rm g/cm}^3$). This idea is based on the 
notion that 
the vortices by means of which the superfluid ``rotates'' can be 
pinned to the crust and therefore be prevented from moving
outwards (and thus spinning the superfluid down) as the 
crust undergoes magnetic braking. Once sufficient strain between
the two components has been built up the vortices undergo 
chaotic ``unpinning'' and a glitch occurs \cite{alpar81}. 
However, very recent observations of possible 
free precession in PSR B1828-11 \cite{stairs} seem to indicate a 
much lower degree of superfluid vortex pinning than is usually 
assumed in the glitch models \cite{jones}. Thus the standard 
glitch model 
would seem to be in trouble and our need for an improved 
understanding of neutron star superfluidity and its 
astrophysical manifestation is amply illustrated.

{\em ``Gravitational-wave asteroseismology''}. --- 
The exciting possibility that  
gravitational waves from pulsating neutron stars may prove to be
detectable, and that a  knowledge of the 
various mode frequencies will provide strong constraints on the 
supranuclear equation of state has been discussed in a series of papers 
\cite{astero1,astero2,BCCS98}.  However, the most recent
estimates suggest that such a  
detection with (say) LIGO~II requires mode-excitation to possibly 
unrealistic amplitudes \cite{astero2}. 
The most promising scenario
corresponds to the various modes being excited following the 
formation of the neutron star after gravitational collapse. 
This could potentially lead to an energy equivalent to 
$10^{-6}M_\odot c^2$ being radiated as gravitational waves, and there 
is no obvious reason why a comparable energy should not be deposited 
in various nonradial modes of oscillation. But there are still
two problems with this 
scenario: Firstly, the event rate would be rather low unless the 
waves from
neutron stars born in the Virgo supercluster (at a distance of
roughly 20~Mpc) can 
be seen (and this seems unlikely \cite{astero2}). 
Secondly, it may be difficult 
to detect the associated gravitational waves even from events in 
our own galaxy should significantly less energy be radiated.
Basically, the recent estimates suggest that the detectability of 
the waves using an advanced LIGO 
detector is likely to be marginal even from the most 
optimistic astrophysical scenarios. On the other hand,  the possibility
that such observations may help shed light on the true nature of
matter at extreme densities provides strong motivation 
for pursuing work in this direction. It also seems likely that
the first  direct detection of gravitational waves will be the 
initial step towards a true revolution in the way that we view the 
Universe, leading to the development of  
detectors with significantly increased sensitivity. We feel that --- 
as a new generation of detectors is about to come on-line ---
it is appropriate to speculate about future challenges for this field.

\begin{figure}[tbh]
\centerline{\epsfxsize=8.5cm \epsfbox{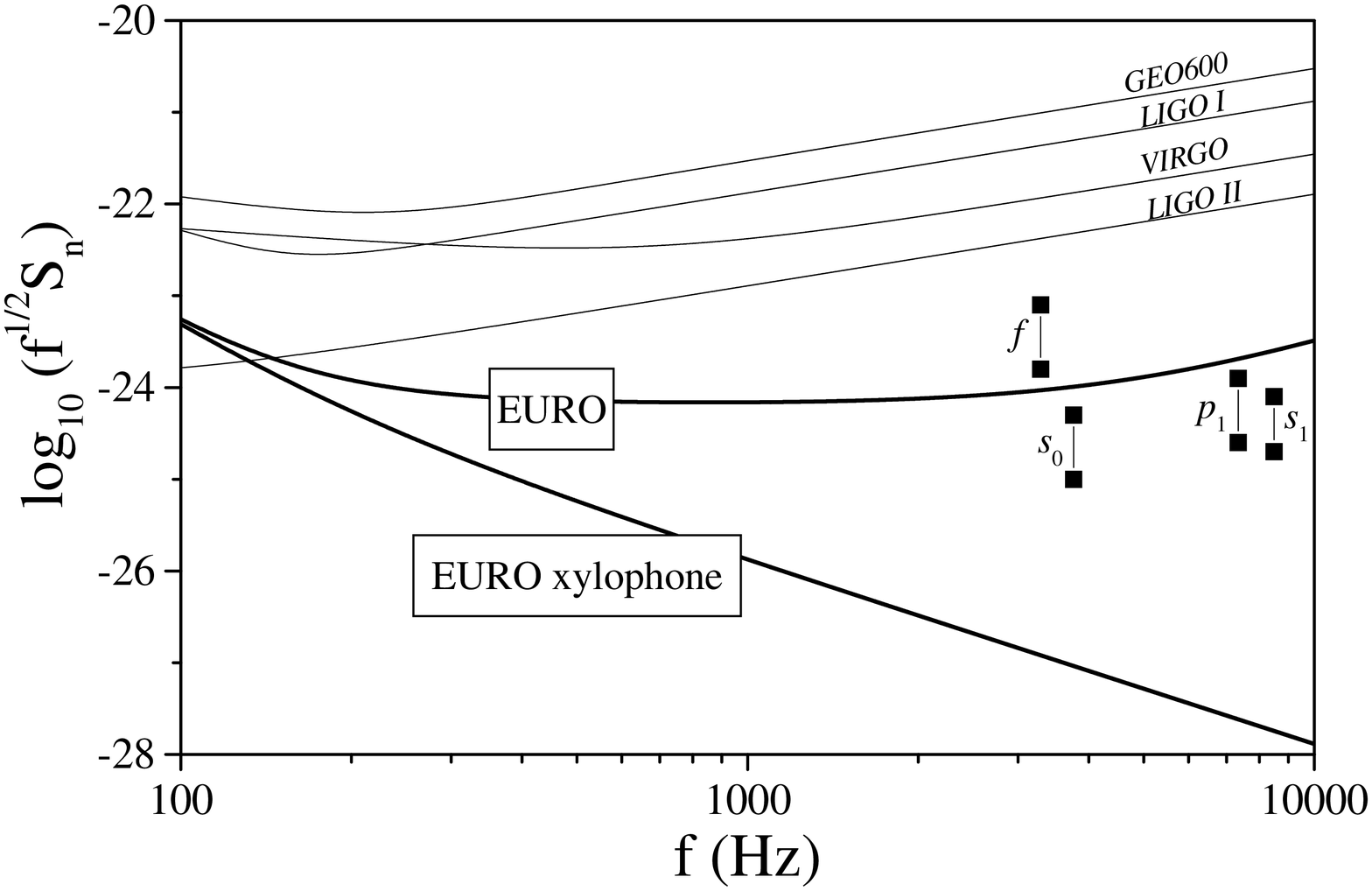}}
\caption{The spectral noise density for the new generation of 
laser-interferometric gravitational-wave detectors that will come 
online in the next few years (thin lines) is compared to the more 
speculative estimates for the EURO detector (solid lines).  A key 
feature of this advanced configuration is that it may operate several 
narrowbanded interferometers as a xylophone, thus reaching high 
sensitivity at khz frequencies. We also indicate the effective 
gravitational-wave amplitudes from the glitch-induced mode-oscillations
discussed in the text.}
\label{figure1}\end{figure}

The various gravitational-wave detector groups are already 
discussing possible technological improvements
that may be achievable in the future. 
As an example of a suitably advanced instrument we will 
take the so-called EURO detector, for which the noise-level has been 
estimated by Sathyaprakash and Schutz 
(for further details see \cite{euro}).  We 
will consider
two possible configurations: In the first, the sensitivity at
high frequencies is limited by the photon shot-noise, while 
the second configuration reaches beyond this limit by running 
several narrowbanded (cryogenic) interferometers as a ``xylophone''.
The corresponding noise-curves are illustrated, and 
compared to the current generation of interferometers, in 
Figure~\ref{figure1}.  

By comparing the results of previous discussions of 
gravitational-wave asteroseismology to the data in Figure~\ref{figure1}
we immediately see that a 
EURO detector would provide a superb
instrument for studying pulsating neutron stars. This means that 
previously suggested strategies \cite{astero2}  
for unveiling the supranuclear equation of 
state may be put to the test. However, 
we want to be more ambitious than this. We also want to 
be able to infer the parameters of neutron star superfluidity and 
possibly
shed light on the mechanism for pulsar glitches.

{\em Superfluid neutron star seismology}. --- An ordinary fluid 
neutron star has a plethora of pulsation modes, the most familiar 
of which are the pressure p-modes (the fundamental of which is known as 
the f-mode) and the gravity g-modes. 
In addition to these families of modes, the non-zero shear modulus 
in the neutron star crust provides support for families of modes, 
the Coriolis force leads to the presence of the
so-called r-modes in a rotating star et cetera (see \cite{vh80} for a
concise description). Basically,  each extra 
piece of physics of  a detailed neutron star model brings (at least) one 
new family of non-radial modes into play. So what happens as 
the core of the star becomes superfluid?

In the simplest description, a superfluid neutron star core can be 
discussed in terms of two distinct fluids. One of these fluids 
represents the superfluid neutrons and the other fluid represents
a charge-neutral conglomerate of the remaining components (which
are expected to be coupled on a 
relatively short timescale). The fact that these two 
fluids --- which we will refer to as the ``neutrons'' and the 
``protons'' --- can flow more or less independently provides one of the 
main distinguishing 
dynamical features of a superfluid neutron star. 
The neutrons and the protons are coupled via the 
so-called entrainment effect, whereby the 
momentum of one constituent carries along part of the mass of the other 
constituent (analogous to the way that the momentum of one of the 
constituents is thought to
carry along (or entrain) some of the mass of the 
other in a mixture of the two superfluids Helium three and 
Helium four \cite{sfluid}).  A superfluid is 
locally irrotational, but it mimics large-scale rotation
by forming a dense array of quantized vortices.
Because of the entrainment effect 
the flow of neutrons around these vortices will 
induce a flow also in a fraction of the protons.  This leads to
magnetic fields being attached to the vortices and because of the 
electromagnetic attraction of the electrons to the protons (the 
timescale of which is very short) there will then be a dissipative 
scattering of the electrons. This dissipative mechanism is known as  
mutual friction, and since it serves to damp out any relative motion 
between the neutrons and the protons it is expected to be relevant
both for models of the spin-evolution of a pulsars \cite{als} and 
discussions of mode-oscillations \cite{lm00}.  
 
However, the crucial parameters of neutron star
superfluidity are not very well determined. For example, 
the  strength of mutual friction depends crucially on the 
ratio between the 
``bare'' and ``effective'' proton masses $m_p$ and $m_p^*$. 
Estimates of this ratio suggest that $0.3 \le  {m_p^* / m_p} \le 0.8$.
Clearly, any observation that could help determine this ratio (and 
constrain other relevant parameters) must be seen as an attractive 
proposition.  This prompts the main question motivating this letter: 
Can gravitational-wave asteroseismology provide us with some 
of the desired answers? 
As a first step towards answering this question we need to understand 
the nature of the pulsation modes
of a superfluid neutron star core, and the effect that entrainment
has on the spectrum of the emerging gravitational waves.

In a recent study \cite{acpuls},
aimed at comparing and contrasting  oscillations of 
normal and superfluid neutron star cores, we have discussed this issue
in detail. Briefly, our main conclusions are as follows:
There are two sets of (predominately acoustic) pulsation modes in a 
superfluid core.  Both mode-families are associated with 
high-frequency modes and the oscillation frequencies 
are interlaced in the spectrum.
One set of modes is the familiar p-modes, for which the two fluids 
tend to move together. The other set of modes 
(exemplified by $s_0$ and $s_1$ in Table~\ref{table1}) is  
distinguished by the fact that the protons and neutrons are largely 
``countermoving''.
The existence of these two classes of modes has been established 
in several studies, but their true nature and the relation to the normal 
fluid case has not been clearly explained previously.
Of particular interest for our current discussion is the fact that
the superfluid set of modes is strongly 
dependent on the entrainment parameters.  From the relevant local 
dispersion 
relation (see \cite{acpuls} for a complete discussion) one finds that
superfluid modes can be qualitatively represented by 
the following, approximate, solution for the local mode-frequency:
\begin{equation}
\omega^2_s \approx {m_{p} \over m^*_p} {l (l + 1) \over r^2} c^2_p \ , 
\label{soundspeed}\end{equation}
where $c^2_p$ is (roughly) the sound speed in the proton fluid, $r$ 
is the radial coordinate, and $l$ is the index of the relevant
spherical harmonic $Y_{lm}(\theta,\varphi)$ used to describe the 
angular dependency of the mode.  From this relation it is clear that 
an observation of these modes would provide potentially unique 
information regarding the nature of large scale superfluidity.

We have recently performed the first ever (fully relativistic) 
calculations of both frequency and damping rate of the oscillation modes
of a superfluid neutron star \cite{acl}. 
 This calculation clearly 
distinguishes  the two families of pulsation modes,
and highlights the fact that the superfluid mode frequencies are strongly
dependent on the parameters of entrainment (mainly the average
value of the effective proton mass $m^*_p$, cf. (\ref{soundspeed})). 
The first few 
mode-frequencies and the associated gravitational-wave damping times
for one of our model stars 
are listed in Table~\ref{table1}. (We note in passing that some authors 
\cite{sed} have suggested that the superfluid modes will not radiate 
gravitational waves.)  The given mode-results were obtained for a 
non-rotating stellar model and a simple equation of state corresponding to 
a combination of two polytropes. These results should provide a good 
``order of magnitude'' insight into future results for realistic 
supranuclear equations of state. In particular, since
most observed pulsars are slowly 
rotating (in the sense that they spin at a fraction of the mass-shedding 
limit) these results should be relevant for all but the fastest 
millisecond pulsars.  Likewise, Mendell \cite{M91} has shown that 
two-fluid (neutron and proton) models of neutron star superfluidity 
(where magnetic fields, vortex pinning effects, etc. are ignored)
are reasonable when the core matter oscillations are in the kilohertz range.

\begin{table}
\caption{The frequency and damping rate for the first few modes of 
Model~II of Comer et al \cite{acl}. 
The corresponding star has mass $1.36M_\odot$ 
and radius 7.9~km, and could be considered as a ``reasonable'' model 
for a superfluid
neutron star core. We also show the gravitational-wave
signal-to-noise ratios resulting from 
the glitch model discussed in the main text. The results correspond to 
an advanced EURO detector with (model 1) and without (model 2) photon
shotnoise, respectively. The 
lower estimate is for a Crab glitch while the upper estimate follows 
from the Vela data. 
\label{table1}}
\begin{ruledtabular}
\begin{tabular*}{\hsize}{lcccr}
Mode & $f$ (kHz) & $t_d$ (s) & Model 1 & Model 2\\
\hline
$f$ & 3.29 & 0.092  & 0.4---6 & 300---$4.7\times10^3$\\
$p_1$ & 7.34 & 1.01 & 0.08---1.2 & 680---$1\times10^4$ \\
\hline
$s_0$ & 3.76 & 15.75 & 0.3---4.8 & 350---$5.4\times10^3$ \\
$s_1$ & 8.49 & 1.29 & 0.06---0.9 & 780---$1.2\times 10^4$\\
\end{tabular*}
\end{ruledtabular}
\end{table}

{\em Unveiling the nature of pulsar glitches}. --- 
To what extent will future gravitational-wave observations  be able to 
detect the various pulsation modes of a superfluid neutron star?  
Let us assume that a typical gravitational-wave signal from a
neutron star pulsation mode takes the
form of a damped sinusoidal, i.e.
\begin{equation}
h(t) = {\cal A} e^{-(t-T)/t_d} \sin [ 2\pi f (t-T)] \quad \mbox{ for } 
       t > T
\end{equation}
where $T$ is the arrival time of the signal at the detector (and
$h(t)=0$ for $t<T$). 
Using standard results for the gravitational-wave flux \cite{astero2},
the amplitude ${\cal A}$ of the signal can be expressed in terms of the 
total energy radiated through the mode:
\begin{equation}
{\cal A} \approx 7.6\times 10^{-24} \sqrt{{\Delta E_\odot \over 
10^{-12} } {1 \mbox{ s} \over t_d}}  
 \left( {1 \mbox{ kpc} \over d } \right) 
\left({ 1 \mbox{ kHz} \over f} \right)  \ .
\end{equation}
where $\Delta E_\odot = \Delta E/M_\odot c^2$.
Finally, the signal-to-noise ratio for this signal can be 
estimated from \cite{astero2}
\begin{equation}
\left({S \over N} \right)^2 = { 4Q^2 \over 1+4Q^2} {{ \cal A}^2 t_d 
\over 2S_n}
\label{sign}\end{equation}
where the ``quality factor'' is $Q=\pi f t_d$ and 
$S_n$ is the spectral noise density of the detector (in 
Figure~\ref{figure1} 
we show the dimensionless strain  $\sqrt{f} S_n$ for various 
detector configurations). 

The main question here 
is: What amount of energy should one assume to be channeled
through the various modes? In the previous studies it was assumed that 
as much as $10^{-5}M_\odot c^2$ could be radiated \cite{astero2}. 
However, this number is likely
only relevant (if at all) for the oscillations of the remnant following 
a strongly asymmetric
gravitational collapse. It is not useful for our present 
considerations since such a nascent neutron star will be hot (above
$10^{10}$~K) enough for the core not to be superfluid.    
For a reasonable scenario, we turn to the indications that young 
neutron stars are seismically active. As a suitably simple
model scenario we will assume that oscillations in the superfluid core 
are excited following a glitch. The released energy
can then be estimated from
\begin{equation}
\Delta E \approx I \Omega \Delta \Omega \approx (10^{-6}-10^{-8}) I 
\Omega^2
\end{equation}
where $\Omega = 2\pi /P$.
In this formula it is appropriate to use the moment of inertia 
$I\sim 10^{45}$~gcm$^2$ 
of the entire star, since the 
spin-up incurred during the glitch remains on timescales that are 
much longer than the estimated coupling timescale between the crust 
and the 
core fluid. 
Using this formula, and the data for the Crab and Vela pulsars (see 
Table~\ref{table2}) we can estimate the energy associated with typical 
glitch events. (We note that these estimates are similar to ones already
in the literature \cite{franco00,hart00}.)

Assume that a comparable amount of energy goes into 
exciting oscillations in the core superfluid. This is obviously ad 
hoc, but it  provides a reasonable order-of-magnitude
starting point for this kind of discussion. Using 
the spectral density
estimated for the EURO detector we then readily estimate the 
associated signal-to-noise ratio from (\ref{sign}). The results of this 
exercise are listed in Tables~\ref{table1}-\ref{table2}. 
From this data we can see 
that the various modes would be marginally detectable given this level 
of excitation and a third generation
detector limited by the photon shotnoise. 
But if this limit can be surpassed by configuring several 
narrowbanded interferometers as a xylophone, the achievable 
signal-to-noise ratio is excellent. 
Besides estimating the signal-to-noise ratio for the various
oscillation modes for a given radiated energy, we can ask related 
questions 
relevant for the inverse problem. For example, we can confirm 
that the oscillation frequencies can be extracted with good accuracy
from the data.  This then enables us to distinguish clearly between the 
``normal fluid'' f and p-modes and the superfluid s-modes.  In other 
words, we would have the information required not only to infer the 
mass and radius of the star \cite{astero2}, we could also hope to 
constrain the
parameters of neutron star superfluidity.  
In addition to providing this important information, these observations 
could provide a unique insight into the glitch mechanism 
itself since the multipolar
structure of the excited modes reflects the symmetry (or lack thereof)
of the triggering mechanism.

 \begin{table}[h]
\caption{Data for archetypal glitching pulsars.
\label{table2}}
\begin{ruledtabular}
\begin{tabular*}{\hsize}{lcccr}
PSR & $P$ (ms) & $d$ (kpc) & $\Delta \Omega / \Omega$ & $\Delta E/M_\odot 
c^2$ \\
\hline
Crab & 33 & 2 & $10^{-8}$ & $2\times10^{-13}$ \\
Vela & 89 & 0.5 & $10^{-6}$ & $3\times10^{-12}$\\
\end{tabular*}
\end{ruledtabular}
\end{table}

{\em Issues}. --- This is a very exciting time for gravitational 
physics. The opening of a 
new window to the Universe may well lead 
to a fundamental change in the way that we view Nature. For example, 
it is generally expected that gravitational-wave observations will 
provide us with crucial information regarding the details of neutron star
physics, eg. by constraining the supranuclear equation of state. 
Such information may to a certain extent be obtained  by analysing 
signals from 
inspiraling neutron star binaries, see for example \cite{vall}. 
As we have argued in this Letter
an observation of the various 
modes of oscillation of (say) a glitching pulsar would 
provide a truly unique probe of the internal physics that could 
help improve our understanding of, in particular, 
large-scale superfluidity. The potential precision of this 
method easily surpasses 
other proposed methods for studying neutron star superfluidity, 
eg.~by observed cooling data \cite{page00}. 
Of course, it seems likely that our proposed scheme requires 
the development of detectors with sensitivity beyond 
the goals set for the advanced LIGO configuration.  
However, it is clear that,
 even though the construction of a gravitational-wave
detector with the sensitivity of EURO provides a serious challenge, 
a successful effort in this direction would be 
richly rewarded. 

If we are to achieve this ambitious goal we must also make 
significant progress on the theoretical side. 
We need dynamical studies of general relativistic 
superfluid neutron stars comprising a crust in the outer parts 
and other possible exotic phases of 
matter in the core. 
We need to improve our understanding of the mechanisms that 
couple the various components of a realistic neutron star, and 
estimate the relevant coupling timescales.
We need numerical simulations that investigate the
extent to which the various oscillation modes are excited by 
plausible mechanisms. These, and related (eg.~regarding rapidly 
spinning stars and possible mode-instabilities 
\cite{ak00}), topics 
are tremendously exciting since they 
challenge our understanding 
of the very extremes of physics.

NA is a Philip Leverhulme Prize fellow supported by PPARC grant 
PPA/G/S/1998/00606.  GLC was 
supported by a Saint Louis University SLU2000 Faculty Research Leave 
Award as well as a visiting fellowship from EPSRC in
the UK via grant number GR/R52169/01.

\end{document}